# Self-sustained oscillations in whiskers without vortex shedding

Shayan Heydari[a,1], Mitra J. Z. Hartmann[b,c], Neelesh A. Patankar[c], and Rajeev K. Jaiman[a]

[a]Department of Mechanical Engineering, The University of British Columbia, Vancouver, BC Canada V6T 1Z4; [b]Department of Biomedical Engineering, Northwestern University, Evanston, IL 60208; [c]Department of Mechanical Engineering, Northwestern University, Evanston, IL 60208

**Sensing the flow of water or air disturbance is critical for the survival of many animals: flow information helps them localize food, mates, and prey and to escape predators. Across species, many flow sensors take the form of long, flexible cantilevers. These cantilevers are known to exhibit sustained oscillations when interacting with fluid flow. In the presence of vortex shedding, the oscillations occur through mechanisms such as wake- or vortex-induced vibrations. There is, however, no clear explanation for the mechanisms governing the sustained oscillation of flexible cantilevers without vortex shedding. In recent work, we showed that a flexible cylindrical cantilever could experience sustained oscillations in its first natural vibration mode in water at Reynolds numbers below the critical Reynolds number of vortex shedding. The oscillations were shown to be driven by a frequency match (synchronization) between the flow frequency and the cantilever's first-mode natural frequency. Here, we use a body-fitted fluid-structure solver based on the Navier-Stokes and nonlinear structural equations to simulate the dynamics of a cantilevered whisker in the air at a subcritical value of Reynolds number. Results show that second-mode synchronization governs the whisker's sustained oscillation. Wavy patterns in the shear layer dominate the whisker's wake during the vibrations, indicating that parallel shear layers synchronize with the whisker's motion. As a result of this synchronization, oval-shaped motion trajectories, with matching streamwise and cross-flow vibration frequencies, are observed along the whisker. The outcomes of this study suggest possible directions for designing artificial bio-inspired flow sensors.**

synchronization | lock-in | subcritical flow | sensing

Many animals expertly use complex, three-dimensional (3D) fluid flow information to navigate and explore their environments. Insects, crustaceans, rodents, and pinnipeds have specialized sensors that enable them to detect and localize flow sources (1–3) and, in some cases, to track fluid currents and wakes (3–6). Notably, across all of these species, the fluid-detecting sensors – hairs, antennae, and whiskers – take the form of long, flexible cantilevers with a bluff cross-section. To understand how animals use these specialized sensors to acquire and exploit complex flow information, and to design artificial devices that can replicate these capabilities, requires a robust understanding of the fluid mechanics involved. However, it is particularly challenging to study fluid flow around such flexible structures because the fluid and the structure form a dynamically coupled system involving a bi-directional feedback effect known as fluid-structure interaction (FSI) (7, 8).

Due to the FSI effect, a long flexible bluff body can exhibit cross-flow vibrations in a steady incident flow normal to its length (9). While a bluff-body section implies the existence of flow separation accompanied by two shear layers bounding a relatively broad wake, a steady incident flow means that there exists no organized transient or oscillatory characteristic in the incident flow. The flow dynamics around a bluff body depend on the Reynolds number, $Re = \rho^f U D / \mu^f$, where $\rho^f$ is the density of the fluid, $U$ is the free stream velocity, $D$ is the characteristic length, i.e., the diameter in the case of a circular cylinder, and $\mu^f$ is the dynamic viscosity of the fluid. A pioneering study of a circular cylinder excited into vibration by a flowing fluid can be attributed to Strouhal (10) and Rayleigh (11). According to the findings of Strouhal and Rayleigh, the frequency $f_{vs}$ of the aeolian tone produced by the relative motion of a thin wire and the airflow varies with the diameter $D$ and with the speed $U$ of the relative motion. The value of the dimensionless group, the so-called Strouhal number $St = f_{vs} D/U$, is dependent upon the Reynolds number $Re$, geometry, the roughness of the body's surface, and the free stream turbulence.

For the case of a circular cylinder at very low $Re$ ($Re \lesssim 4$), the fluid wraps around the cylinder with boundary layers fully attached. For $4 \lesssim Re \lesssim 45$, however, the flow separation appears behind the cylinder, where symmetric re-circulation bubbles are formed. Further increase of $Re$ results in unsteady vortices and the flow bifurcates to a time-periodic state where opposite-signed vortices are periodically shed from the cylinder surface (12). This periodic vortex shedding generates oscillatory forces exerting on the cylinder and could result in large-amplitude cross-flow oscillations. Since such oscilla-

> **Significance Statement**
>
> Several species rely on the fluid flow information collected by their sensory organs to explore and navigate their surroundings. These organs resemble long, flexible cantilevers that exhibit sustained oscillations when exposed to fluid currents. Although much is known about how flexible cantilevers oscillate under fluid flow, specifically periodic vortex shedding, little is known about the mechanisms governing the oscillatory motion of these structures without the influence of vortex shedding. This work employs a high-fidelity numerical solver to demonstrate that synchronization is a general phenomenon that underlies the sustained oscillation of flexible cantilevers independent of vortex shedding. The implemented numerical framework overcomes the challenges posed by conducting detailed physical experiments and formulating analytical models and is desirable in developing flow sensors.

R.K.J., N.A.P., and M.J.Z.H. designed research; S.H. performed research and analyzed data; S.H., M.J.Z.H., and R.K.J. wrote the paper; S.H. performed all simulations; M.J.Z.H. contributed to data analysis; R.K.J., M.J.Z.H., and N.A.P. advised S.H. on all aspects of the work; R.K.J. secured the funding for the work, interpreted data, and offered a vision for the work.

The authors declare no competing interest.

[1]To whom correspondence should be addressed. E-mail: sheydari@mail.ubc.ca

tions are associated with vortex shedding, they are termed vortex-induced vibrations (VIVs). Within the field of FSI, significant research has been devoted to characterizing the oscillatory dynamics of cylindrical structures during VIVs. Specifically, for flow conditions that translate into Reynolds numbers above the critical Reynolds number of vortex shedding, i.e., $Re > Re_{cr} \approx 45$ (12, 13), VIVs have been shown to occur for a certain range of fluid and structural parameters, within which the vortex shedding frequency, $f_{vs}$, deviates from the Strouhal's relationship (vortex shedding frequency of a stationary cylinder) and becomes equal or close to the cylinder's natural frequency (9, 14, 15).

Although the oscillatory motion of cylindrical structures under vortex shedding is adequately described in the literature, there is no satisfactory explanation for the mechanism underlying the sustained oscillation of these structures at Reynolds numbers below the critical Reynolds number of vortex-shedding. Specifically, the mechanism governing the quasi-periodic motion of a rat's slender whisker (2, 16) in the subcritical regime of $Re$ is still unknown. To begin to understand how flexible bluff bodies, such as whiskers, exhibit sustained oscillations in the subcritical regime of $Re$, we recently analyzed the FSI of a flexible cylindrical cantilever in water for $Re \in [20, 40]$ (17). Results of our numerical experiments showed that the interplay between viscous dissipation and flow energy (which may or may not be in the form of discrete vortices), can generate oscillatory motions in the cylinder for $Re$ as low as 22. We found that synchronization (sometimes referred to as lock-in) between the fluid loading frequency and the cylinder's first-mode natural frequency is the mechanism that drives the oscillations for $Re \geq 22$ in water. It was this synchronization behavior that was shown to result in periodic vortex-shedding downstream. In other words, we showed that in the subcritical regime of $Re$ in water, periodic vortex-shedding occurs as a consequence of synchronization.

The results obtained in our previous work lead to a view in which synchronization is regarded as a general phenomenon (18) that could give rise to various forms of flow-induced vibrations, including VIVs, fluttering, and galloping (see Discussion). As such, we proposed that VIVs, fluttering, and galloping could all be viewed as manifestations of synchronization, and not as separate physical phenomena. This view helps generalize our understanding of the coupling mechanisms in fluid-structure systems and offers a unified explanation for the observed sustained oscillations in flexible cylinders without vortex shedding. As further validation of this conjecture, in the present study, we extend our analysis to the FSI of a long, flexible, cylindrical cantilever in the air at the subcritical regime of $Re$. Due to the highly complex nonlinear multi-physics involved and the difficulty of running experiments at such low Reynolds numbers, a detailed analysis of this problem is exceedingly challenging via physical experiments and analytical techniques. Hence, we implement a high-fidelity 3D numerical framework to investigate the coupled dynamics of the cantilever and pinpoint the mechanism governing its oscillatory response. All results have significant implications for the study of biologically-based flow sensing and engineered fluid sensors.

## Results

The present investigation is based on a model of the "C2" rat whisker (Fig. 1*A*), with a length to base diameter ratio of 230.2 (2). The whisker's geometry is taken as a cylinder of length $L$ and a constant circular cross-section of diameter $D$ (Fig. 1*B*). Although real rat whiskers taper from base to tip and have an intrinsic curvature (19), these geometric features are neglected in the present study. We expect taper and curvature to change the underlying eigenmode distribution along the whisker, but not affect the fundamental results.

A three-dimensional computational domain (Fig. 1*C*) has been constructed to simulate how the whisker responds to the incoming airflow. Details of the computational grid and convergence studies are provided in *SI Appendix, Finite-element mesh convergence and verification*. We study the whisker's response for five values of airflow speeds, $U$, between 0.2-1 (m/s). The Reynolds number is set equal to 40 and the mass ratio parameter, which corresponds to the ratio of the density of a real rat whisker to the density of air, is equal to 1000 (19).

The primary goal of the simulations is to answer two specific questions. First, are sustained oscillations observed along the whisker in the subcritical regime of $Re$, and, second, what is the mechanism driving the oscillations at this $Re$ regime? Although we do not aim to perform an extensive parameter sweep, several simulation results, including the whisker's vibration frequency and motion trajectory, could be compared with experimental observations, identified in Fig. 1*D*.

**Synchronization at the $2^{nd}$ natural mode explains the response characteristics of the whisker.** The motion trajectory of the whisker at $Re = 40$ and $U = 1$ (m/s) is provided in Fig. 2*A*. When uniform flow passes over the whisker, a sustained oscillatory response takes place after some time has passed and the whisker exhibits oval-shaped motion trajectories. A zero-displacement node is present at a normalized distance of $z/L \approx 0.8$ (from the fixed end of the whisker). The existence of the node is also evident from the scalograms of the whisker's response (Figs. 2*B* and 2*C*). According to the scalogram plots, the whisker's motion exhibits standing wave patterns in both the streamwise and cross-flow directions. Note that the scale of the cross-flow response is an order of magnitude larger than the streamwise response.

To investigate the mechanism underlying the sustained oscillations, we considered all of the forces acting at each point along the whisker. Using the coordinate system in Fig. 2*A*, the cross-sectional "drag" force is in the streamwise direction ($\pm$ x), while the cross-sectional "lift" force is in the cross-flow direction ($\pm$ y). The shear force, which acts in the $\pm$ z-direction, can be neglected since it will be nearly constant with time along the entire length (17). The power spectra of the drag and lift coefficients are compared with the power spectra of the dimensionless streamwise (x-direction) and cross-flow (y-direction) vibration amplitudes in Figs. 2*D* and 2*E*. Figure 2*D* reveals a clear match between the frequency of the drag coefficient and the frequency of the streamwise oscillation at a value of 236.3 (Hz). There is a similar match between the frequency of the lift coefficient and the frequency of the cross-flow vibration also at a value of 236.3 (Hz) (Figure 2*E*).

For a real C2 whisker, experimental measurements suggest a second-mode natural frequency of $200 - 225$ (Hz) (2), with the theoretical value of the frequency being approximately equal to 208.3 (Hz) (20). For our current study, the theoretical value of the second-mode natural frequency is 236.73 (Hz) (see *SI Appendix, Natural frequency and mode shape calculation*), which matches well with the values obtained using numerical simulations. Figures 2*F* and 2*G* compare the snapshots of the whisker's profile in the streamwise and cross-flow directions obtained using the coupled numerical solver with the expected profile of an uncoupled Euler-Bernoulli beam in its second natural mode of vibration. Based on the match between the whisker's profile and the profile of the Euler-Bernoulli beam, we conclude that the whisker's response is predominantly in its second natural mode. Hence, the second-mode lock-in/synchronization is considered the intrinsic preferred frequency during the coupling between the whisker and the airflow at the subcritical regime of $Re = 40$.

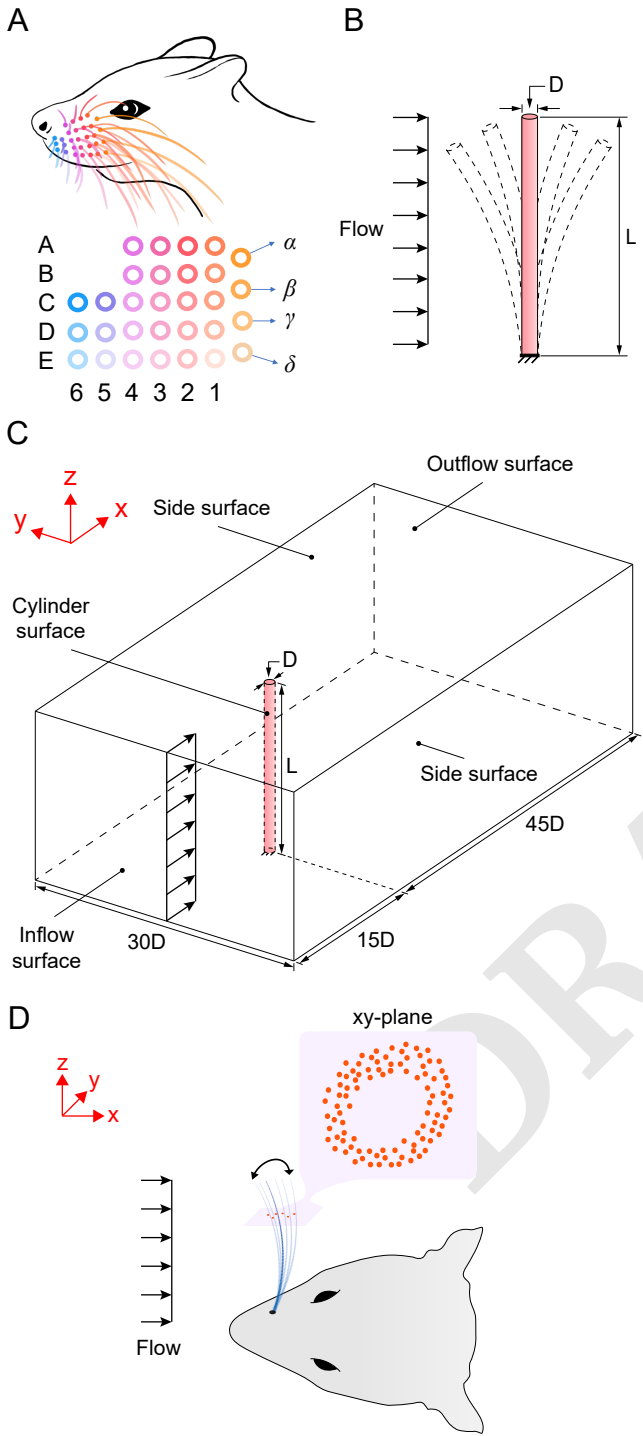

**Fig. 1.** (*A*) Whiskers are arranged in regular rows and columns on the rat's face and identified by their row (letter) and column (number) identity. We focus on a cylindrical model of the C2 whisker. (*B*) Rat whisker modeled as a flexible cantilever cylinder of length $L$ and constant circular cross-section of diameter $D$. (*C*) Three-dimensional computational domain. The whisker is positioned at an offset distance of $15D$ and $45D$ from the inflow and outflow surfaces, respectively. Fixed structural support is imposed at one end of the whisker, and the no-slip boundary condition is applied at the cylinder surface. A uniform flow is given at the inflow surface, and the symmetry boundary condition is applied to the side surfaces. For the outflow surface, the traction-free boundary condition is specified. (*D*) Although the goal of the present study is not to perform a comprehensive parameter sweep, we qualitatively compare the results to experimental observations relating to the vibration frequencies and ellipse-like motion of the whisker's tip.

**Synchronization occurs over a broad range of speeds and explains the oval-shaped tip trajectories observed experimentally.** We next investigated the range of air speeds between 0.2-1 (m/s) to explain the oval-shaped tip trajectories observed in experiments (16). Figure 2*H* plots the variations of the root-mean-square (RMS) value of the dimensionless streamwise and cross-flow vibration amplitudes over the range of studied air speeds. According to Fig. 2*H*, the whisker exhibits sustained oscillations for $U \geq 0.4$ (m/s). The amplitude of the streamwise oscillation has a small value (less than two percent of the whisker's diameter) and remains approximately the same as the flow speed increases. However, the amplitude of the cross-flow vibration increases monotonically from $0.07D$ at $U = 0.4$ (m/s) to approximately $0.32D$ at $U = 1$ (m/s). This trend is consistent with the behavior of a typical flutter/galloping response, in which the vibration amplitude also increases when the flow speed is increased (9, 21).

Figure 3*A* illustrates the tip motion trajectories for the values of studied air speeds. It is seen that the whisker's tip exhibits oval-shaped motion trajectories for $U \geq 0.4$ (m/s). The observed oval trajectories are explained by analyzing the power spectra of the oscillations. Results show that the dominant frequency of the oscillations in the streamwise and cross-flow directions is approximately 236.3 (Hz) for $U \in [0.4, 1]$. Due to the equality of the vibration frequencies, the whisker's motion follows oval trajectories such that in every cycle of its motion, the whisker completes one period in both directions.

**Shear layer oscillations in the whisker's wake are consequences of synchronization.** Although the previous sections have described the mechanism by which the oscillations are sustained, they have not yet identified the origin of the oscillations. To investigate their origin, we examined the wake region behind the whisker at $U = 0.8$ (m/s). Figure 3*B* shows an isometric view of the z-vorticity contours along the full length of the whisker, and expanded x-y plane views for select values of $z/L$. At the locations $z/L = 0$ and $z/L \approx 0.8$, i.e., regions with zero displacements, the wake is steady and symmetric. However, in regions where the whisker vibrates, an unsteady wake, with oscillating parallel shear layers, is present (Fig. 3*B*). Importantly, despite the whisker's large-amplitude vibration, its wake does not exhibit periodic vortex-shedding.

These results stand in distinct contrast to the wake of a flexible cylinder in the water. According to our previous study (17), the wake of a long, flexible cantilever in water

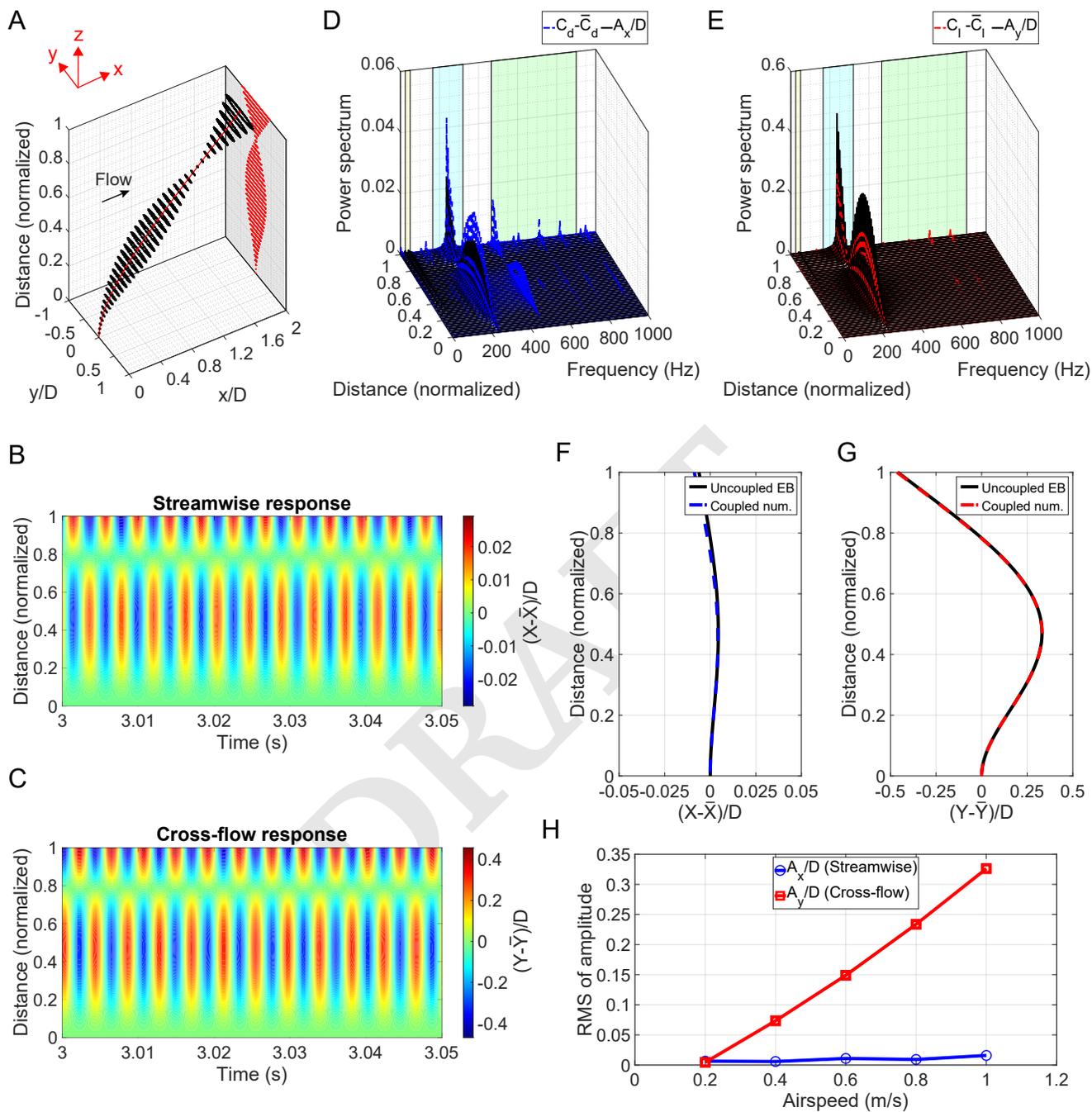

**Fig. 2.** Synchronization as the mechanism of self-sustained oscillations in whiskers without vortex shedding. For plots (*A*)-(*G*), the distance is normalized with respect to the cylinder length $L$ and results are obtained for $U = 1$ (m/s). All plots correspond to $Re = 40$ and mass ratio of 1000. (*A*) Motion trajectories of the whisker (black lines). The red-filled dots, connected with a red line, represent the mean position of the whisker, and the red line projections into the y-z plane show snapshots of the whisker's profile in the cross-flow direction. (*B-C*) Scalograms reveal the presence of standing waves in both streamwise and cross-flow directions; if traveling waves were present, there would be changes with respect to time on the x-axis of the plots. (*D-E*) Power spectra of $C_d - \overline{C_d}$ and $A_x/D$ alongside the power spectra of $C_l - \overline{C_l}$ and $A_y/D$. The bar sign ( ¯ ) corresponds to the mean value of the parameter. The force coefficients, $C_d$ and $C_l$, are equal to the drag and lift forces, respectively, divided by $0.5\rho^f U^2 DL$. The regions highlighted in yellow, cyan, and green indicate the widest possible range for a real rat whisker's first-, second-, and third-mode natural frequencies, respectively (2). (*F-G*) Comparisons between instantaneous profiles of the whisker in the (*F*) streamwise and (*G*) cross-flow directions obtained using numerical simulations with the expected profile of an uncoupled Euler-Bernoulli (EB) beam in its second natural vibration mode. The numerical results are captured at time $t = 3.3$ (s). (*H*) RMS value of the streamwise and cross-flow vibration amplitudes.

**4** | www.pnas.org/cgi/doi/10.1073/pnas.XXXXXXXXXX    Heydari *et al.*

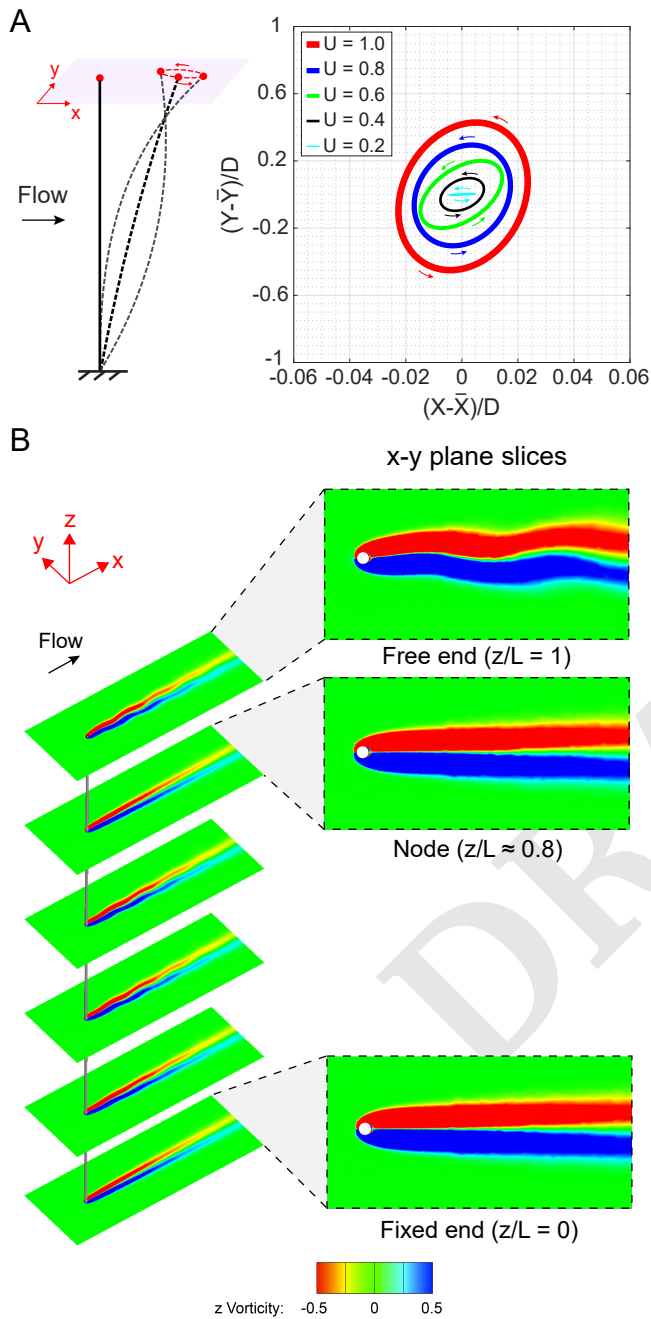

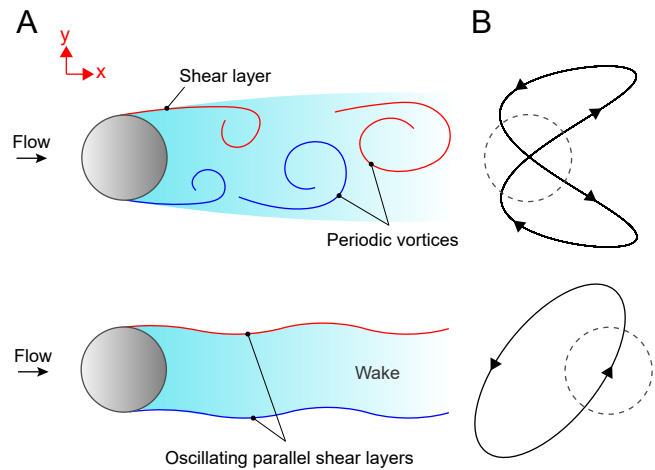

**Fig. 4.** (*A*) Schematic representation of the wake region behind the whisker during synchronization at $Re = 40$; (top) Motion of the whisker in water leads to the formation of vortices and their shedding to the downstream; (bottom) Motion of the whisker in the air causes oscillating parallel shear layers. (*B*) Synchronization of the whisker in the subcritical regime of $Re$ results in (top) figure-eight shaped motion trajectories in water and (bottom) oval-shaped motion trajectories in air. Arrows indicate the direction of motion.

**Fig. 3.** (*A*) (left) The undeformed whisker (black solid line) bends in the streamwise direction first (black dashed line) and then vibrates in its second natural mode (dashed gray lines). The x-y plane at the free end of the whisker is shown in pink. The red filled dots show the position of the probing node on the x-y plane; (right) Representative oval-shaped motion trajectories at the tip. (*B*) Isometric view of the x-y plane z-vorticity contours alongside the x-y plane view of the contours at the free end ($z/L = 1$), zero-displacement node ($z/L \approx 0.8$), and the fixed end ($z/L = 0$) of the whisker at $U = 0.8$ (m/s).

exhibits periodic vortex-shedding patterns during synchronization for the same $Re = 40$. The presence of periodic wake vortices in the subcritical regime of $Re$ in water has also been observed for the case of a freely-vibrating rigid cylinder under similar conditions (22, 23). Our current results also contrast with results for a stationary or non-vibrating cylinder; in these classic cases, the cylinder's wake is steady and symmetric for $Re < Re_{cr}$ (12, 13, 24). Thus, the wake of a flexible cylindrical cantilever during synchronization in air resembles neither the wake of the cantilever in water nor the wake of a stationary cylinder.

What is the origin of the shear layer oscillations? For the flexible cantilever in air, several eigenmodes and natural frequencies are excited, and the second mode lock-in/synchronization arises as an intrinsic preferred frequency. The amplitude of the oscillation increases as the airflow speed increases, similar to a flutter/galloping. At the same time, the system is devoid of vortex shedding. Hence, shear layer oscillations are the consequences – not the origin – of the second-mode synchronization. In other words, synchronization leads to the vibration of the cantilever along with the oscillating parallel shear layers.

## Discussion

In this work, we examined the FSI of a whisker model using numerical simulations. We showed that the whisker oscillates in its second natural mode while interacting with low-speed airflow. Here, we elaborate on the wake dynamics and discuss the differences between the whisker's motion trajectory in water and air. Figure 4*A* illustrates the wake region behind the whisker in water and air at $Re = 40$. For the whisker in water, we have previously shown that synchronization leads to periodic vortex shedding in the wake region (17). However, wake vortices are not formed behind the whisker in the air. Instead, oscillating parallel shear layers are observed during the vibrations (Fig. 4*A*). This difference in the wake features makes the oscillatory response of the whisker in the air quite different from its response in water. For the whisker in water, synchronization generates figure-eight shaped motion trajectories along the whisker (Fig. 4*B*), where the vibrations are in the first natural vibration mode (17). However, for the whisker in the air, oval-shaped trajectories are observed across the whisker, and the vibrations are shown to be in the second natural vibration mode. In both water and air, the motion of the whisker locks onto the motion of the wake; hence, its vibration frequency is linked to the frequency of the wake. For the whisker in water, the shedding of vortices downstream results in the frequency of the streamwise oscillation being twice that of the cross-flow vibrations (17). However, for the whisker in air, since the wake region is devoid of vortex shedding, both the streamwise and cross-flow vibrations have equal frequencies, linked to the frequency of the oscillating parallel shear layers. The differences in the ratio of the vibration frequencies lead to figure-eight shaped motion trajectories in water and oval-shaped trajectories in air.

Based on the results of this work and our previous study (17), we conclude that the existence of discrete wake vortices is not necessary to excite the vibrations during synchronization. We imply that synchronization is an intrinsic characteristic of a fluid-structure system that can give rise to periodic vortex shedding or oscillating parallel shear layers in the subcritical *Re* regime. This viewpoint helps classify VIVs and movement-induced instabilities, such as flutter/galloping, as manifestations of synchronization and not as separate physical phenomena. For instance, a typical VIV response could be regarded as a 1:1 synchronization between the vortex-shedding frequency and the structure's first-mode natural frequency (14). Similarly, fluttering could be considered as a 1:1 synchronization between two structural frequencies (21, 25), such as torsional and bending mode frequencies (26), while galloping could be seen as a 1:1 synchronization between a fluid flow frequency, which differs from the vortex-shedding frequency, and one of the structure's natural frequencies. In all these instances, the vibrations are excited by the fluid flow energy and sustained through the synchronization mechanism.

## Methods

Here, we briefly describe the employed continuum mechanics framework for simulating the coupled dynamics of the whisker in a flowing fluid. The numerical methodology closely follows that outlined in Ref. (27). Detailed information about the governing equations is provided in *SI Appendix, Governing equations and details of the numerical framework*. We use three-dimensional incompressible Navier-Stokes equations, in the moving boundary Arbitrary Lagrangian-Eulerian (ALE) framework, coupled with the nonlinear hyperelastic structural equation to examine the dynamics of the whisker. An iterative partitioned fluid-structure coupling algorithm transfers the information between the fluid and structure domains at the body-fitted fluid-structure interface. A stabilized Petrov-Galerkin finite element method is employed to discretize the fluid domain $\Omega^f$ into $n_{el}$ non-overlapping finite elements $\Omega^e$ in space such that $\Omega^f = \bigcup_{e=1}^{n_{el}} \Omega^e$. The variational generalized-$\alpha$ time integration method (28) is adopted for the temporal discretization of the fluid variables.

We solve the structural equations in the Lagrangian coordinate system with the fluid spatial points moving according to the ALE moving mesh framework. To model the dynamically deforming structure, we consider a nonlinear hyperelastic model as described in (29). The system of structural equations is solved using the standard Galerkin finite element technique by means of isoparametric elements for curved boundaries. A three-dimensional common-refinement method has been used to transfer data between the incompressible fluid and the nonlinear hyperelastic structure for conditions when there is a mismatch between the fluid and structure meshes at the fluid-structure interface. Further information concerning the accuracy of the fluid-structure coupling strategy and suitability of the solver for modeling the coupled dynamics of the whisker could be found in Refs. (27, 30).

**ACKNOWLEDGMENTS.** The authors would like to acknowledge the Natural Sciences and Engineering Research Council of Canada (NSERC) for funding the project. The research was partly enabled through computational resources and services provided by the Digital Research Alliance of Canada, and the Advanced Research Computing facility at the University of British Columbia.

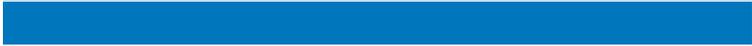

## Supporting Information for

**Self-sustained oscillations in whiskers without vortex shedding**

**Shayan Heydari, Mitra J. Z. Hartmann, Neelesh A. Patankar, and Rajeev K. Jaiman**

**Shayan Heydari**
**E-mail: sheydari@mail.ubc.ca**

**This PDF file includes:**

Supporting text
Fig. S1
Tables S1 to S3
SI References

**Supporting Information Text**

**Governing equations and details of the numerical framework**

For the high-fidelity modeling of continuum fluid-structure interactions, we consider fully coupled nonlinear partial differential equations for the fluid and the structural systems. The governing equations for the fluid are applied in an arbitrary Lagrangian-Eulerian form while the dynamical structural equation is formulated in a Lagrangian way, and the interface conditions are enforced between the two physical fields. For the sake of completeness, we provide a brief review of our numerical fluid-structure framework.

**Incompressible Navier-Stokes equations.** First, we describe the governing equations for the incompressible flow at the continuum level. We consider a three-dimensional spatial fluid domain $\Omega^f(t) \in \mathbb{R}^3$ with a piecewise smooth boundary $\Gamma^f(t)$ at time $t$. Two types of boundary conditions are considered for the boundaries of the fluid domain; the non-interface Neumann boundary $\Gamma^f_N(t)$, and the Dirichlet boundary $\Gamma^f_D(t)$. The Navier-Stokes equations for an incompressible viscous fluid flow with the boundary conditions are given as

$$\rho^f \frac{\partial \boldsymbol{u}^f}{\partial t} + \rho^f \boldsymbol{u}^f \cdot \boldsymbol{\nabla} \boldsymbol{u}^f = \boldsymbol{\nabla} \cdot \boldsymbol{\sigma}^f + \rho^f \boldsymbol{b}^f \quad \text{on} \quad \Omega^f(t) , \tag{1}$$

$$\boldsymbol{\nabla} \cdot \boldsymbol{u}^f = 0 \quad \text{on} \quad \Omega^f(t) , \tag{2}$$

$$\boldsymbol{u}^f = \boldsymbol{u}^f_D \quad \forall \boldsymbol{x}^f \in \Gamma^f_D(t) , \tag{3}$$

$$\boldsymbol{\sigma}^f \cdot \boldsymbol{n}^f = \boldsymbol{h}^f \quad \forall \boldsymbol{x}^f \in \Gamma^f_N(t) , \tag{4}$$

$$\boldsymbol{u}^f = \boldsymbol{u}^f_0 \quad \text{on} \quad \Omega^f(0) , \tag{5}$$

where $\rho^f$ represents the fluid density and $\boldsymbol{u}^f = \boldsymbol{u}^f(\boldsymbol{x}^f, t)$ is the fluid velocity at each spatial point $\boldsymbol{x}^f \in \Omega^f(t)$ and time $t$. The body force per unit mass acting on the fluid is denoted by $\boldsymbol{b}^f$. The Dirichlet condition given in Eq. (3) imposes fluid velocity $\boldsymbol{u}^f_D$ on the Dirichlet boundary $\Gamma^f_D(t)$. The Neumann boundary condition is given by $\boldsymbol{\sigma}^f \cdot \boldsymbol{n}^f = \boldsymbol{h}^f$, where $\boldsymbol{n}^f$ is the unit normal vector to the Neumann boundary $\Gamma^f_N(t)$. The initial velocity field in the fluid domain at $t=0$ is given by $\boldsymbol{u}^f_0$. The Cauchy stress tensor for a Newtonian fluid is represented by $\boldsymbol{\sigma}^f$ and is defined as

$$\boldsymbol{\sigma}^f = -p\boldsymbol{I} + \mu^f \left( \boldsymbol{\nabla} \boldsymbol{u}^f + \left( \boldsymbol{\nabla} \boldsymbol{u}^f \right)^T \right) , \tag{6}$$

where $p$ is the pressure and $\mu^f$ is the dynamic viscosity of the fluid flow. To couple the fluid equations formulated in the Eulerian coordinate system and the structural equations given in the Lagrangian coordinate system, the Arbitrary Lagrangian-Eulerian (ALE) coordinate system is introduced to the fluid equations. The ALE coordinate system combines the advantages of both the Eulerian and Lagrangian descriptions by expressing the equations in the new reference coordinate $\boldsymbol{\chi}$. The one-to-one mapping from the node $\boldsymbol{\chi}$ in the referential coordinate system to its corresponding spatial point $\boldsymbol{x}^f$ in the Eulerian coordinate system is written as

$$\boldsymbol{x}^f = \boldsymbol{\Phi}^f(\boldsymbol{\chi}, t) \qquad \forall t, \tag{7}$$

where $\boldsymbol{\Phi}^f$ is the mapping function. The velocity with which the spatial coordinates move with respect to the referential coordinate system $\boldsymbol{\chi}$ is called the mesh velocity $\boldsymbol{u}^m$ and is defined as

$$\boldsymbol{u}^m = \left. \frac{\partial \boldsymbol{x}^f}{\partial t} \right|_{\boldsymbol{\chi}} . \tag{8}$$

The material or total derivative of an arbitrary function $f(\boldsymbol{\chi}, t)$ in the ALE coordinate system can be written as

$$\frac{Df}{Dt} = \left. \frac{\partial f(\boldsymbol{\chi}, t)}{\partial t} \right|_{\boldsymbol{\chi}} + \frac{\partial f(\boldsymbol{x}^f, t)}{\partial \boldsymbol{x}^f} \left. \frac{\partial \boldsymbol{x}^f}{\partial \boldsymbol{\chi}} \frac{\partial \boldsymbol{\chi}}{\partial t} \right|_{\boldsymbol{x}^s}, \tag{9}$$

where $\boldsymbol{x}^s$ is a material point at time instant $t$. Considering $f = \boldsymbol{x}^f$, the material derivative of the spatial coordinate $\boldsymbol{x}^f$ is given as

$$\boldsymbol{u}^f = \frac{D\boldsymbol{x}^f}{Dt} = \left. \frac{\partial \boldsymbol{x}^f}{\partial t} \right|_{\boldsymbol{\chi}} + \frac{\partial \boldsymbol{x}^f}{\partial \boldsymbol{\chi}} \left. \frac{\partial \boldsymbol{\chi}}{\partial t} \right|_{\boldsymbol{x}^s} . \tag{10}$$

Substituting Eq. (10) and Eq. (8) into Eq. (9), the material derivative of the function $f$ is rewritten into the following form

$$\frac{Df}{Dt} = \left. \frac{\partial f(\boldsymbol{\chi}, t)}{\partial t} \right|_{\boldsymbol{\chi}} + (\boldsymbol{u}^f - \boldsymbol{u}^m) \frac{\partial f(\boldsymbol{x}^f, t)}{\partial \boldsymbol{x}^f} . \tag{11}$$

Hence, the incompressible Navier-Stokes equations on $\Omega^f(t)$ in the ALE framework are derived from Eqs. (1), (2), and (11)

$$\rho^f \left. \frac{\partial \boldsymbol{u}^f}{\partial t} \right|_{\boldsymbol{\chi}} + \rho^f (\boldsymbol{u}^f - \boldsymbol{u}^m) \cdot \boldsymbol{\nabla} \boldsymbol{u}^f = \boldsymbol{\nabla} \cdot \boldsymbol{\sigma}^f + \rho^f \boldsymbol{b}^f, \tag{12}$$

$$\boldsymbol{\nabla} \cdot \boldsymbol{u}^f = 0. \tag{13}$$

**Nonlinear hyperelastic structure.** We solve the structural equations in the Lagrangian coordinate system with the fluid spatial points moving according to the ALE moving mesh framework. The continuous differential equation for the structure, along with the boundary conditions, are presented in this subsection. Let us consider a three-dimensional structural domain $\Omega^s(0) \in \mathbb{R}^3$ with material coordinate $\boldsymbol{x}^s$ at time $t = 0$. For a coupled fluid-structure system, the structural domain includes a piecewise smooth boundary $\Gamma^s(t)$ which consists of the non-interface Neumman boundary $\Gamma_N^s(t)$, the Dirichlet boundary $\Gamma_D^s(t)$, and the fluid-structure interface boundary $\Gamma^{fs}(t)$. A one-to-one mapping function $\boldsymbol{\varphi}^s(\boldsymbol{x}^s, t)$ is defined to map the reference coordinates of the structure in the initial configuration $\Omega^s(0)$ from points $\boldsymbol{x}^s$ at $t = 0$ to their position in the deformed state $\Omega^s(t)$ at time instant $t$. Considering $\boldsymbol{d}^s(\boldsymbol{x}^s, t)$ as the structural displacement, the mapping function is written as

$$\boldsymbol{\varphi}^s(\boldsymbol{x}^s, t) = \boldsymbol{x}^s + \boldsymbol{d}^s(\boldsymbol{x}^s, t). \qquad [14]$$

Thus, the structural velocity $\boldsymbol{u}^s$ and acceleration $\boldsymbol{a}^s$ are defined as

$$\boldsymbol{u}^s = \frac{\partial \boldsymbol{\varphi}^s}{\partial t}, \qquad \boldsymbol{a}^s = \frac{\partial \boldsymbol{u}^s}{\partial t} = \frac{\partial^2 \boldsymbol{\varphi}^s}{\partial t^2} = \frac{\partial^2 \boldsymbol{d}^s}{\partial t^2}. \qquad [15]$$

The governing structural equations in their most general form can be written as

$$\rho^s \frac{\partial^2 \boldsymbol{\varphi}^s}{\partial t^2} = \boldsymbol{\nabla} \cdot \boldsymbol{\sigma}^s + \rho^s \boldsymbol{b}^s \quad \text{on} \quad \Omega^s(t), \qquad [16]$$

$$\boldsymbol{u}^s = \boldsymbol{u}_D^s \quad \forall \boldsymbol{x}^s \in \Gamma_D^s(t), \qquad [17]$$

$$\boldsymbol{\sigma}^s \cdot \boldsymbol{n}^s = \boldsymbol{h}^s \quad \forall \boldsymbol{x}^s \in \Gamma_N^s(t), \qquad [18]$$

$$\boldsymbol{\varphi}^s = \boldsymbol{\varphi}_0^s \quad \text{on} \quad \Omega^s(0), \qquad [19]$$

$$\boldsymbol{u}^s = \boldsymbol{u}_0^s \quad \text{on} \quad \Omega^s(0), \qquad [20]$$

where $\rho^s$, $\boldsymbol{\sigma}^s$, and $\boldsymbol{b}^s$ denote the structural density, the stress tensor, and the body force per unit mass acting on the structure, respectively. The Dirichlet condition on the structural velocity and the Neumann condition on the stress tensor are denoted by $\boldsymbol{u}_D^s$ and $\boldsymbol{h}^s$, respectively. The unit normal vector to the Neumann boundary $\Gamma_N^s$ is shown by $\boldsymbol{n}^s$. Equation (19) is the initial condition for the position of the structure, while Eq. (20) represents the initial condition for the structural velocity.

To model the dynamically deforming structure, we consider a nonlinear hyperelastic model. Using the principle of virtual work, the variational form of Eq. (16) for the model is written as

$$\int_{\Omega^s(0)} \rho^s \partial_t u_i^s \delta v_i^s \mathrm{d}\Omega + \int_{\Omega^s(0)} J^s \sigma_{ij}^s \delta L_{ij}^s \mathrm{d}\Omega$$

$$- \int_{\Omega^s(0)} \rho^s b_i^s \delta v_i^s \mathrm{d}\Omega - \int_{\Gamma_N^s(0)} h_i^s \delta v_i^s \eta^s \mathrm{d}\Gamma = 0. \qquad [21]$$

In the above equation, $u_i^s$ denotes the structural velocity in the deformed state; $\delta v_i^s = \delta v_i^s(\boldsymbol{x}^s)$ is the virtual velocity field which satisfies $\delta v_i^s = 0$ along the Dirichlet boundary $\Gamma_D^s$, $\delta L_{ij}^s = \partial \delta v_i^s / \partial \varphi_j^s$ is the gradient of the virtual velocity field, $J^s = \det(\boldsymbol{F}^s)$ is the Jacobian of the deformation gradient tensor $\boldsymbol{F}^s$ with $F_{ij}^s = \delta_{ij} + \frac{\partial d_i^s}{\partial x_j}$, and $\sigma_{ij}^s$ is the Cauchy stress. The inverse surface Jacobian $\eta^s$ is used to map the boundary surface of $\Omega^s(t)$ back to the reference configuration $\Omega^s(0)$ by $\eta_{ij}^s = \partial x_i^s / \partial \zeta_j$, where $\zeta_j$ is the isoparametric coordinate. The Cauchy stress $\sigma_{ij}^s$ is related to the left Cauchy–Green stress via the neo-Hookean constitutive law as

$$\sigma_{ij}^s = \frac{\mu^s}{(J^s)^{5/3}}(B_{ij}^s - \frac{1}{3}B_{kk}^s \delta_{ij}) + K^s(J^s - 1)\delta_{ij}, \qquad [22]$$

where $\mu^s$ and $K^s$ are the shear modulus and the bulk modulus of the structure, respectively, and $B_{ij}^s$ is the left Cauchy–Green stress given by $B_{ij}^s = F_{ik}^s F_{jk}^s$.

**Treatment of fluid-structure interface.** We need to satisfy the continuity of velocity and traction at the fluid-structure interface. Let $\Gamma^{fs} = \partial \Omega^f(0) \cap \partial \Omega^s$ be the fluid-structure interface at time $t = 0$ and $\Gamma^{fs}(t) = \boldsymbol{\varphi}^s(\Gamma^{fs}, t)$ be the interface at time $t$. The required conditions to be satisfied are

$$\boldsymbol{u}^f(\boldsymbol{\varphi}^s(\boldsymbol{x}^s, t), t) = \boldsymbol{u}^s(\boldsymbol{x}^s, t), \qquad [23]$$

$$\int_{\boldsymbol{\varphi}^s(\gamma,t)} \boldsymbol{\sigma}^f(\boldsymbol{x}^f, t) \cdot \boldsymbol{n}^f \mathrm{d}\Gamma + \int_\gamma \boldsymbol{\sigma}^s(\boldsymbol{x}^s, t) \cdot \boldsymbol{n}^s \mathrm{d}\Gamma = 0, \qquad [24]$$

where $\boldsymbol{\varphi}^s$ is the position vector that maps the reference coordinates of the structure in the initial configuration $\Omega^s(0)$ from points $\boldsymbol{x}^s$ to their position in the deformed state $\Omega^s(t)$ at time instant $t$, $\boldsymbol{u}^s = \partial \boldsymbol{\varphi}^s / \partial t$ is the velocity of the structure at time $t$, $\boldsymbol{n}^f$ and $\boldsymbol{n}^s$ are the unit normal vectors to the interface surfaces in the fluid and structural domains, respectively, $\gamma$ is any part of the fluid-structure interface $\Gamma^{fs}$ in the reference configuration, and $\boldsymbol{\varphi}^s(\gamma, t)$ is the corresponding fluid part at time $t$. The above conditions are satisfied such that the fluid velocity is exactly equal to the velocity of the structure, i.e., continuity of velocities, and the motion of the structure is governed by the fluid forces at the fluid-structure interface, i.e., continuity of tractions. The temporal discretization of both the fluid and the structural equations is embedded in the generalized-$\alpha$ framework by employing the classical Newmark approximations in time.

**Mesh motion equation.** To account for the changes in the location of the fluid-structure interface, the motion of each spatial point is explicitly controlled in the fluid domain, while the kinematic consistency of the discretized interface is satisfied. The spatial points $\boldsymbol{x}^f \in \Omega^f(t)$ are moved within the fluid domain by solving the equation

$$\nabla \cdot \boldsymbol{\sigma}^m = 0, \qquad [25]$$

where $\boldsymbol{\sigma}^m$ is the stress experienced by the spatial points due to the strain induced by the interface deformation. Assuming that the fluid mesh behaves as an elastic material, $\boldsymbol{\sigma}^m$ could be expressed as

$$\boldsymbol{\sigma}^m = (1 + k_m) \left[ \nabla \boldsymbol{d}^f(\boldsymbol{\chi}, t) + \left( \nabla \boldsymbol{d}^f(\boldsymbol{\chi}, t) \right)^T + \left( \nabla \cdot \boldsymbol{d}^f(\boldsymbol{\chi}, t) \right) \boldsymbol{I} \right], \qquad [26]$$

where $k_m$ is the local element-level mesh stiffness parameter chosen as a function of the element sizes. Equation (25) solves for the fluid mesh nodal displacement $\boldsymbol{d}^f$ that satisfies the boundary conditions

$$\boldsymbol{d}^f(\boldsymbol{x}^s, t) = \boldsymbol{\varphi}^s(\boldsymbol{x}^s, t) - \boldsymbol{x}^s \qquad \text{on} \quad \Gamma^{fs},$$
$$\boldsymbol{d}^f(\boldsymbol{\chi}, t) = 0 \qquad \text{on} \quad \partial\Omega^f(0) \setminus \Gamma^{fs}. \qquad [27]$$

**Partitioned fluid-structure coupling.** The employed partitioned approach involves the following steps at the nonlinear iteration $k$ of the algorithm in the time domain $t \in [t^n, t^{n+1}]$:

1. Solve the nonlinear hyperelastic structural equation for the predictor structural displacement $\boldsymbol{d}^{s,n+1}_{(k+1)}$ based on the fluid forces evaluated at the previous time step $\boldsymbol{f}^{s,n}_k$.

2. Transfer the structural displacement at the fluid-structure interface $\Gamma^{fs}$ to the fluid side by satisfying the ALE mesh compatibility and the velocity continuity conditions at the fluid-structure interface.

3. Solve the mesh motion equation for the displacements of the fluid nodes and calculate the mesh velocity field;

4. Get the updated fluid velocity $\boldsymbol{u}^{f,n+1}_{(k+1)}$ and pressure $p^{n+1}_{(k+1)}$ and evaluate the fluid forces.

5. Transfer the corrected fluid forces to the structural solver to satisfy the dynamic equilibrium at the fluid-structure interface.

When the algorithm achieves convergence, the solver advances in time after updating the variable values at time instant $t^{n+1}$.

## Finite-element mesh convergence and verification

**Fluid solver.** The employed computational domain is discretized into unstructured hexahedral finite element grids with a boundary layer mesh around the structure. To ascertain the accuracy of the formulation and its implementation, computations are carried out for flow past a stationary cylinder at the Reynolds number of $Re = 60$. Three sets of meshes with an increasing number of elements are selected for this study. The finest mesh, denoted by M3, is taken as the reference case. The results of the study in terms of the root-mean-square (RMS) value of the lift coefficient $C_L^{rms}$ and the Strouhal number $St = f_{vs}D/U$, where $f_{vs}$ is the vortex-shedding frequency, $D$ is the cylinder's diameter, and $U$ is the flow velocity, are provided in Table S1. According to Table S1, the computed values of $C_L^{RMS}$ and $St$ for the M2 and M3 meshes are in good agreement with the reported experimental (1) and numerical (2) values. Hence, we conclude that the M2 mesh is adequate to compute the flow field around the cylinder.

**Structural solver.** A standalone convergence study has also been done to ensure the accuracy of the results for the nonlinear hyperelastic structural solver. For this purpose, we have examined the deformation of a flexible cylindrical cantilever under a uniform distributed load $q_0$. The parameters used for this study are summarized in Table S2. The results in terms of the maximum displacement, denoted by $\delta_{max}$, are gathered for three sets of structural meshes with an increasing number of elements and compared with the value of $\delta_{max}$ obtained from the Euler-Bernoulli beam theory (3). Table S3 summarizes the grid convergence study results. We conclude that the M3 mesh has achieved sufficient convergence; therefore, it is used as the reference case. As seen in Table S3, the computed value of $\delta_{max}$ is in good agreement with the solution of the Euler-Bernoulli beam theory.

## Structural dynamics using linear Euler-Bernoulli beam equation

As part of our analysis, we investigated the fluid-structure interaction of the whisker using the linear Euler-Bernoulli beam equation coupled with incompressible Navier-Stokes equations. For the structural solver, we considered a mode superposition approach, where the displacements from the mean position of the whisker were assumed to be small and characterized based on the normal vibration modes found using an eigenvalue analysis. The implementation details of this approach can be found in Ref. (4). We studied the FSI of the whisker in the air at $Re = 40$, and airflow speed of 0.8 (m/s). The results in terms of the time history of the streamwise and cross-flow tip displacements are provided in Fig. S1. As seen in Fig. S1, the whisker's vibratory response dies out after some time has passed. These results suggest that weakly nonlinear effects play a role in exciting the whisker's oscillatory modes and need to be considered when modeling the FSI of the whisker.

**Natural frequency and mode shape calculation**

The i$^{\text{th}}$ mode natural frequency of a flexible cantilever in a vacuum is given by

$$f_i = \frac{\lambda_i^2}{2\pi L^2}\sqrt{\frac{EI}{m}}, \quad [28]$$

where $i$ is the mode number, $E$ is Young's modulus, $I$ is the second moment of area, $m$ is the mass per unit length given by $m = \rho^s \pi D^2/4$, where $\rho^s$ is the cantilever's density, and $\lambda_i$ is the dimensionless frequency parameter for the $i^{th}$ mode of vibration. The values for $\lambda_i$ could be found in Ref. (5). Considering $\lambda_2 = 4.6941$ (5), $E = 6$ GPa, $\rho^s = 1320$ kg/m$^3$, $L = 34.3$ mm, and $D = 0.149$ mm, the second-mode natural frequency of the cantilever studied in this work is calculated as $f_2 = 236.73$ (Hz).

The mode shapes of the cantilever are written as the sums of sine, cosine, sinh, and cosh functions of $\lambda_i z/L$ such that

$$\begin{aligned} S_i(z) &= \cosh\left(\frac{\lambda_i z}{L}\right) - \cos\left(\frac{\lambda_i z}{L}\right) - \sigma_i \sinh\left(\frac{\lambda_i z}{L}\right) \\ &\quad + \sigma_i \sin\left(\frac{\lambda_i z}{L}\right), \end{aligned} \quad [29]$$

where $S_i$ is the shape associated with the $i^{th}$ mode of vibration, $z$ is the distance from the fixed end of the cantilever, and $\sigma_i$ is the non-dimensional parameter dependent on the mode number. For the second mode of vibration, $\sigma_2 = 1.0185$ (5), hence, the corresponding mode shape is calculated as

$$\begin{aligned} S_2(z) &= \cosh\left(\frac{4.6941 z}{L}\right) - \cos\left(\frac{4.6941 z}{L}\right) \\ &\quad - 1.0185 \sinh\left(\frac{4.6941 z}{L}\right) + 1.0185 \sin\left(\frac{4.6941 z}{L}\right). \end{aligned} \quad [30]$$

Considering $S_2(z) = 0$ yields the non-zero value of $z = 0.7834 L$ as the position of the zero-displacement node.

**Table S1. Mesh convergence, validation, and verification results for the stationary rigid cylinder problem. The percentage differences are the relative errors compared to the results for the M3 mesh. A constant time-step size $\Delta t = 0.001$ is employed.**

| Mesh | Fluid elements | $C_L^{RMS}$ | $St$ |
|---|---|---|---|
| M1 | 13677 | 0.096 (4.00 %) | 0.132 (5.71%) |
| M2 | 27357 | 0.1 | 0.138 (1.43%) |
| M3 | 54714 | 0.1 | 0.140 |
| Experimental (1) | - | 0.1 | 0.142 |
| Numerical (2) | 20731 | 0.1 | 0.137 |

**Shayan Heydari, Mitra J. Z. Hartmann, Neelesh A. Patankar, and Rajeev K. Jaiman**

**Table S2. Parameters for the problem of the flexible cylindrical cantilever under a uniform distributed load.**

| Parameters | Value (SI) |
|---|---|
| Cylinder diameter, $D$ | 0.149 |
| Cylinder length, $L$ | 34.3 |
| Structure's density, $\rho^s$ | 1320 |
| Young's modulus, $E$ | $3 \times 10^9$ |
| Distributed loading per unit length, $q_0$ | 0.075 |

**Table S3. Grid convergence study results for the flexible cylindrical cantilever under a uniform distributed load. The values inside the parentheses represent the relative error compared to the results of the M3 mesh.**

| Mesh        | Structural elements | $\delta_{max}$   |
|-------------|---------------------|------------------|
| M1          | 15200               | 0.134 (21.17 %)  |
| M2          | 30400               | 0.163 (4.12%)    |
| M3          | 60800               | 0.170            |
| Beam theory | -                   | 0.179            |

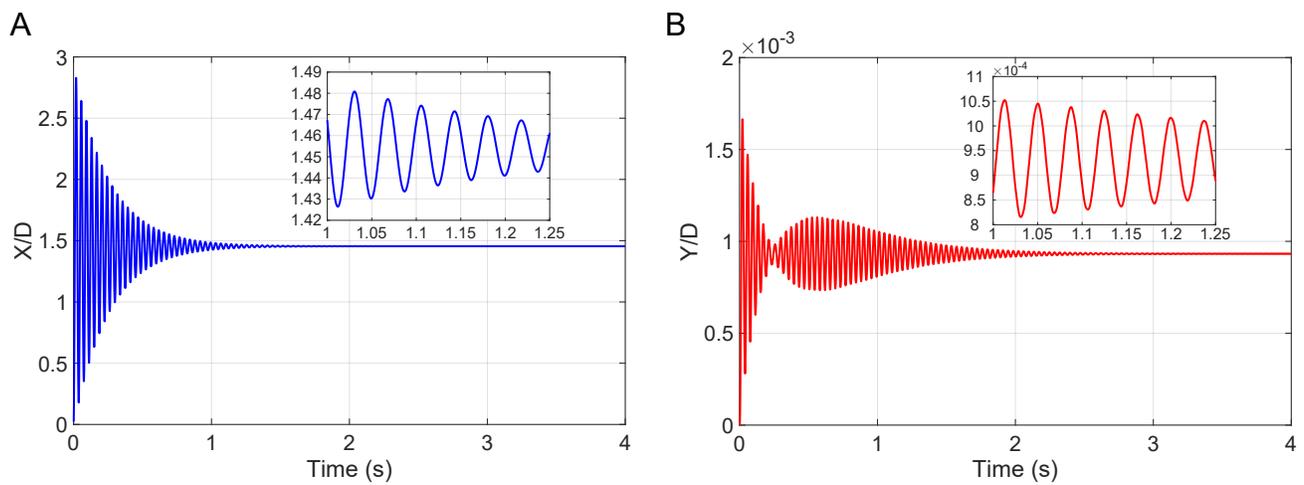

**Fig. S1.** Time history of the whisker's tip displacement in (*A*) streamwise and (*B*) cross-flow directions using the linear Euler-Bernoulli beam equation coupled with incompressible Navier-Stokes equations.

Shayan Heydari, Mitra J. Z. Hartmann, Neelesh A. Patankar, and Rajeev K. Jaiman 9 of 10